\documentclass[apj]{emulateapj}
\usepackage{amsmath}                
\usepackage{amsfonts}               
\usepackage{amssymb}                
\usepackage{epsfig}                 
\usepackage[para,online,flushleft]{threeparttable}

\newcommand{\s}{{~\rm s}}

\newcommand{\G}{{~\rm G}}

\newcommand{\erg}{{~\rm erg}}

\newcommand{\nar}{{~\rm New Astronomy Reviews}}
\newcommand{\na}{{~\rm New Astronomy}}

\begin{document}

\title{Magnetar-powered superluminous supernovae must first be exploded by jets}

\author{Noam Soker\altaffilmark{1} \& Avishai Gilkis\altaffilmark{2}}

\altaffiltext{1}{Department of Physics, Technion -- Israel Institute of Technology, Haifa
32000, Israel; soker@physics.technion.ac.il}
\altaffiltext{2}{Institute of Astronomy, University of Cambridge, Madingley Rise, Cambridge, CB3 0HA, UK; agilkis@ast.cam.ac.uk}

\begin{abstract}
We analyze recent magnetar light-curve modeling of 38 hydrogen-poor superluminous supernovae (SLSNe), and find that the energies of the explosions themselves, that take place before the magnetar energy is released, are more than what the neutrino-driven explosion mechanism can supply for about half of the systems. These SLSNe must have been exploded by a different process than the delayed neutrino mechanism, most likely the jet feedback mechanism (JFM). The conclusion for magnetar modeling of SLSNe is that jets launched at magnetar birth cannot be ignored, not at the explosion itself and not later when mass fall-back might occur. More generally, the present analysis strengthens the call for a paradigm shift from neutrino-driven to jet-driven explosion models of all core collapse supernovae.
\end{abstract}

\section{Introduction}
\label{sec:intro}

Super luminous supernovae (SLSNe) are supernovae that are much brighter at maximum and radiate a much larger energy than typical core collapse supernovae (CCSNe), sometimes over long rise and decline time scales (e.g., \citealt{GalYam2012}). Their peak luminosity is about $10^{44} \erg \s^{-1}$, and the total radiated energy is $\ga 10^{50} \erg$ (e.g., \citealt{Wangetal2016, Arcavietal2016, Sorokinaetal2016, LiuModjaz2017, DeCiaetal2017, Lunnanetal2017}). In many recent papers the extra energy of SLSNe and the long duration of some of them are attributed to energy released by magnetars, i.e., rapidly rotating magnetized neutron stars (e.g., \citealt{Greiner et al2015, Metzgeretal2015, Kangasetal2017, Kasenetal2016, Chenetal2017a, Marguttietal2017, Mazzalietal2017, Metzgeretal2017, Nicholletal2017a, Villaretal2017, Yuetal2017}, to list some works from the last three years).

In recent papers one of us argued that supernovae that are powered at late times by magnetars are most likely exploded by jets \citep{Soker2016Mag, Soker2017Mag2}. In general, the formation of a magnetar requires the pre-collapse core of the stellar progenitor to spin at a high rate. As such a core collapses it forms an accretion disk around the newly born NS or black hole (e.g., \citealt{Gilkis2018}), and jets are likely to be launched (e.g., \citealt{Nishimura2015}). In a recent study \cite{Chenetal2017a} performed 2D simulations of magnetar-powered CCSNe driven by jets. 
In some cases the energy carried by the jets is larger than what is stored in the newly born magnetar. This further suggests that some SLSNe are powered by late jets, as part of the jet feedback mechanism (JFM; e.g., \citealt{Gilkisetal2016}), rather than by, or in addition to, a magnetar.

Many studies over the years mentioned the possible role of jets in exploding CCSNe (.e.g, \citealt{Wheeleretal2002}). Examples from recent years include the axisymmetrical explosions of SN~2015bn \citep{Inserraetal2016}, SN~2013EJ \citep{Mauerhanetal2017}, and of SN~2009ip \citep{Reilletal2017}, as well as asymmetrical CCSN remnants (e.g., \citealt{Lopezetal2014, Milisavljevic2013}). As well, many studies simulated jets in CCSNe (e.g., \citealt{BrombergTchekhovskoy2016, Barnesetal2017, Chenetal2017a}). However, the majority of earlier studies take jets to play significant roles only in rare types of CCSNe. \cite{Sobacchietal2017} speculate that relativistic jets power all Type Ib/c CCSNe. We, on the other hand, strongly support the jet feedback explosion mechanism, according to which \textit{ all CCSNe are exploded by jets} that act in a negative feedback mechanism (e.g., \citealt{PapishSoker2011, GilkisSoker2015, BearSoker2017, Bearetal2017, GrichenerSoker2017, Soker2017RAA}; see \citealt{Soker2016Rev} for a review).

In a recent study \cite{Nicholletal2017b} model the multicolour light curves of 38 hydrogen-poor SLSNe with a magnetar model, and estimate the magnetar and ejecta properties. They take the explosion itself to be driven by neutrinos and have an energy of approximately $10^{51} \erg$. In the present paper we examine the implications of their modeling. In section \ref{sec:sample} we describe the sample of 21 SLSNe we take from their list of 38 SLSNe, and derive the required explosion energies. In section \ref{sec:energy} we discuss our finding that the explosion energy in many of these SLSNe is much above what the neutrino mechanism can supply, and compare with a theoretical prediction of the JFM. In section \ref{sec:rprocess} we present our view that the jets make a very small amount of r-process elements, hence the presence of jets in most (all) CCSNe does not contradict observations. 
In section \ref{sec:summary} we conclude by further strengthening our call for a paradigm shift from neutrino-driven to jet-driven explosion models of all CCSNe.

\section{The explosion energy}
\label{sec:sample}

Our aim is not to refit a magnetar model to each one of the SLSNe, but rather to use the same assumptions and parameters as derived by \cite{Nicholletal2017b} in their modeling, and from that to estimate the energy of the explosion itself, $E_{\rm SN}$.
We use the same equations as given by them.

The rate of energy loss by the magnetar is given by
\begin{equation}
\frac {d E_{\rm mag}}{dt} = - \frac {E_{\rm mag,0}}{t_{\rm mag,0}}
\left(1+ \frac{t}{t_{\rm mag,0}}\right)^{-2},
\label{eq:dotEmag}
\end{equation}
where subscript zero means that the value is taken at $t=0$. The magnetar rotational energy is given by
\begin{equation}
E_{\rm mag}  = 2.6\times 10^{52}
\left(  \frac{M_{\rm NS}}{1.4 M_\odot} \right)^{3/2}
\left(  \frac{P}{1 {\rm ms}} \right)^{-2} \erg,
\label{eq:Emag0}
\end{equation}
and its spin-down time is given by
\begin{equation}
t_{\rm mag}  = 1.5
\left(  \frac{M_{\rm NS}}{1.4 M_\odot} \right)^{3/2}
\left(  \frac{P}{1 {\rm ms}} \right)^{2}
\left(  \frac{B_\perp}{10^{14} {\rm G}} \right)^{-2} {\rm~ day},
\label{eq:tmag}
\end{equation}
where $P$ and $M_{\rm NS}$ are the spin period and mass of the NS, respectively, and ${B_\perp}$ is the component of the magnetic field perpendicular to the spin axis.

The energy of the magnetar at time $t$ is given by
\begin{equation}
E_{\rm mag} \left(t\right) = E_{\rm mag,0} \left(1+\frac{t}{t_{\rm mag,0}} \right)^{-1}.
\label{eq:Emag}
\end{equation}
Since \cite{Nicholletal2017b} estimate the minimum kinetic energy 15 days after bolometric maximum, $E_{\rm k,min} = E_{\rm k} (t_{\rm max}+15)$, we estimate the explosion energy $E_{\rm SN}$ from the relation
\begin{multline}
E_{\rm SN} \simeq E_{\rm k}\left(t_{\rm max}+15 \right) + E_{\rm rad} \left(t_{\rm max}+15 \right) \\
- \left[ E_{\rm mag,0} - E_{\rm mag} \left(t_{\rm max}+15 \right) \right].
\label{eq:ESN}
\end{multline}

There are some uncertainties in the radiated energy. 
The energy radiated up until 15 days after maximum bolometric light, $E_{\rm rad} (t_{\rm max}+15)$, should include also the thermal energy of the ejecta that has not been radiated yet. This is not much. As well, in the JFM we might expect asymmetrical explosions, such that the radiated energy has no spherical symmetry as well. In any case, as the radiated energy is generally smaller than the kinetic energy, these uncertainties are small. For the radiated energy up to 15 days after maximum we take here $50\%$ of the estimated radiated energy we find in the literature, which is usually a minimum limit on the total radiated energy, $E_{\rm rad} (t_{\rm max}+15) =0.5 E_{\rm rad,min}$. 

The kinetic energy of the ejecta, $E_{\rm k}$, is highly uncertain. \cite{Nicholletal2017b} estimate the minimum kinetic energy of the ejecta, $E_{\rm k,min}$, by taking the expansion velocity of the ejecta 15 days after maximum bolometric light, but state that the kinetic energy can be twice as large. For that, we take the kinetic energy of the ejecta 15 days after maximum light to be
\begin{equation}
E_{\rm k}\left(t_{\rm max}+15\right) = \eta E_{\rm k,min},
\label{eq:Ek}
\end{equation}
with $1 \le \eta \le 2$, where $\eta = 1$ represents the \cite{Nicholletal2017b} estimate.

Our criterion to include an object from the listed 38 SLSNe of \cite{Nicholletal2017b} is that the spin-down time at $t=0$ obeys the relation $t_{\rm mag,0} > 0.5 t_{\rm max}$, where $t_{\rm max}$ is the time from explosion to maximum bolometric light. The systems that we do not include in our sample require a more careful and self-consistent treatment. The explosion energy of some of the systems that we do not include in our analysis can in principle be accounted for by the delayed neutrino mechanism, although we consider it unlikely.

We list the names of the 21 SLSNe that satisfy the criterion of $t_{\rm mag,0} > 0.5 t_{\rm max}$, and the values of the quantities we use for our calculations in the first seven columns of Table \ref{Tab:Table1}.
\cite{Nicholletal2017b} list the initial values (at $t=0$) of their modeling for $M_{\rm NS}$, $P$, ${B_\perp}$, and for the minimum kinetic energy 15 days after maximum bolometric light $E_{\rm k,min}$. The time to maximum light we take either from \cite{Nicholletal2017b} or from \cite{DeCiaetal2017}. We take the minimum radiated energy from several sources as indicated in the seventh column.  
 In the last three columns we list the values of the explosion energies for $\eta=1$, $\eta=1.5$, and $\eta=2$, according to equation (\ref{eq:ESN}). 
\begin{table*}[]
\centering 
\caption{SLSN parameters}
\begin{threeparttable}
\begin{tabular}{lccccccccc}
\hline
SLSN &P&B&$M_{\rm NS}$ &$E_{\rm k,min}$ &$t_{\rm max}$&$E_{\rm rad}$&$E_{\rm sn}(1)$&$E_{\rm sn}(1.5)$
              &$E_{\rm sn}(2)$ \\
  &(ms)&($10^{14} \G$)&($M_\odot$)& (foe) & (day) &(foe)&(foe)&(foe)& (foe) \\
\hline
GAIA16apd   & 2.93 & 1.23 & 1.83 & 3.69  & 24      & 1.6$^{\rm K}$&1.1&2.9&4.8 \\ 
PTF12dam    & 2.28 & 0.18 & 1.83 & 3.03  & 57$^{D}$& 1.6$^{\rm D}$&2.6&4.1&5.6 \\ 
SN2015bn    & 2.16 & 0.31 & 1.78 & 3.45  & 72      & 2.3$^{\rm N}$&1.0&2.7&4.4 \\ 
SN2007bi    & 3.92 & 0.35 & 1.81 & 2.37  & 45      & 1.5$^{\rm G}$&2.7&3.9&5.1 \\ 
SN2010gx    & 3.66 & 0.59 & 1.79 & 3.78  & 12      & 1.5$^{\rm P}$  &6.1&8.0&9.9 \\ 
LSQ14mo     & 4.97 & 1.01 & 1.85 & 2.43  & 29      & 0.3$^{\rm C}$&1.9&3.1&4.3 \\ 
PTF09cnd    & 1.46 & 0.1  & 1.82 & 3.29  & 46$^{D}$& 2.0$^{\rm D}$&2.2&3.9&5.5 \\ 
iPTF13ehe   & 2.57 & 0.2  & 1.87 & 4.48  & 75      & 0.4$^{\rm D}$&3.5&5.8&8.0 \\ 
PTF09cwl    & 1.74 & 0.27 & 1.86 & 6.78  & 37$^{D}$& 1.6$^{\rm D}$&2.9&6.3&9.7 \\ 
SN2006oz    & 2.70 & 0.32 & 1.80 & 2.66  & 70      & 0.4$^{\rm E}$&1.0&2.4&3.7 \\ 
PTF09atu    & 1.59 & 0.09 & 1.88 & 8.30  & 50      & 1.6$^{\rm D}$&7.8&11.9&16.1 \\ 
PS1-14bj    & 2.82 & 0.13 & 1.85 & 4.61  &128$^{L}$& 0.8$^{\rm L}$&4.4&6.7&9.0 \\ 
PS1-11ap    & 3.66 & 0.82 & 1.87 & 1.73  & 58$^{L}$& 1.0$^{\rm L}$&0.4&1.3&2.1 \\ 
DES14X3taz  & 2.41 & 0.39 & 1.87 & 5.87  & 55      & 2$^{\rm S}$  &3.8&6.8&9.7 \\ 
PS1-10bzj   & 5.21 & 1.63 & 1.86 & 2.32  & 29$^{L}$& 0.4$^{\rm L}$&1.5&2.7&3.9 \\ 
DES13S2cmm  & 6.59 & 0.73 & 1.76 & 2.31  & 32      & 1$^{\rm A}$  &2.6&3.8&4.9 \\ 
PS1-10ahf   & 2.35 & 0.17 & 1.85 & 4.10  &131      & 1$^{\rm M}$  &2.8&4.9&6.9 \\ 
SCP-06F6    & 1.78 & 0.16 & 1.75 & 8.35  & 88      & 1.7$^{\rm Q}$&5.9&10.1&14.3 \\
PS1-10pm    & 1.31 & 0.06 & 1.85 & 9.76  & 49      & 0.8$^{\rm L}$&8.9&13.7&18.6 \\ 
SNLS-07D2bv & 3.49 & 0.26 & 1.80 & 1.85  & 44      & 0.5$^{\rm H}$&1.7&2.6&3.5 \\ 
SNLS-06D4eu & 3.55 & 0.79 & 1.88 & 3.63  & 44      & 0.6$^{\rm H}$&2.1&4.0&5.8 \\ 
\hline
\end{tabular}
\footnotesize
\begin{tablenotes}
The list of 21 SLSNe analyzed here. The first column lists the name of the SLSNe, followed by $P$, $B_\perp$, and $M_{\rm NS}$ that are the initial rotational period, the component of the magnetic field perpendicular to the spin axis, and the mass of the neutron star. $E_{\rm k,min}$ is the minimum kinetic energy of the ejecta. These four quantities for each SLSN are taken from the modeling of \cite{Nicholletal2017b}. We take the time to maximum light either from \cite{DeCiaetal2017} (those marked by superscript D), or from \cite{Nicholletal2017b}; those that are marked with superscript L also have a maximum time in the study by \cite{Lunnanetal2017}. In the seventh column we list the  minimum radiated energy that we take from one of the aforementioned papers or from one of the following papers: K: \cite{Kangasetal2017}; 
N: \cite{Nicholletal2016ApJ826}; G: \cite{GalYametal2009};
P: \cite{Pastorelloetal2010};  C: \cite{Chenetal2017}; E: \cite{Leloudasetal2012}; 
S: \cite{Smithetal2016}; A: \cite{Papadopoulosetal2015};  M: \cite{McCrumetal2015};
Q: \cite{Quimbyetal2011}; H: \cite{Howelletal2013}.  
The last three columns list the explosion energy as calculated by equation (\ref{eq:ESN}), and for $\eta=1$, 1.5 and 2, respectively. The energy units in the table are foe (fifty one erg), which equals $10^{51} \erg$.
The uncertainties in the values of the parameters that \cite{Nicholletal2017b} derive introduce uncertainties in our derived values of $E_{\rm SN}$, about $50 \%$, that are smaller than the uncertainties that the unknown values of $\eta$ introduce, and hence are not listed here. 
\end{tablenotes}
\end{threeparttable}
\label{Tab:Table1}
\end{table*}

The uncertainties in the values of the parameters that \cite{Nicholletal2017b} derive are large, typically about $20 \% -50 \%$. These imply uncertainties also in our derived value of $E_{\rm SN}$ of about $50 \%$.  We do not list the uncertainties because they are smaller than the uncertainties in the value of $\eta$, that change the value of $E_{\rm SN}$ by up to a factor of about 4 (see Table \ref{Tab:Table1}). As well, adding the uncertainties will mask the values we present here and will not change at all our main conclusion that the delayed neutrino explosion mechanism cannot account for the explosion of the SLSNe we study here. But the uncertainties should be kept in mind for individual objects.


\section{Implications to the jet feedback explosion mechanism}
\label{sec:energy}

The problems of the delayed neutrino mechanism (e.g., \citealt{Papishetal2015a, Kushnir2015}) suggest that it cannot account for even typical CCSNe. In any case, even its supporters agree that the delayed neutrino mechanism cannot account for CCSN explosion energies of $E_{\rm SN} \ga 2 \times 10^{51} \erg$ (e.g., \citealt{Fryer2006, Fryeretal2012, Sukhboldetal2016, SukhboldWoosley2016}). For $\eta=1$ in equation (\ref{eq:Ek}), that is, when the kinetic energy is equal to the minimum kinetic energy estimated by \cite{Nicholletal2017b}, 14 SLSNe have explosion energies the delayed neutrino mechanism cannot account for. This number becomes 20 for the more likely value of $\eta=1.5$.

The explosion energies $E_{\rm SN}(\eta)$ that we estimate in section  \ref{sec:sample} and list in the last three columns of Table \ref{Tab:Table1} are uncertain. First, there is the question of the kinetic energy that \cite{Nicholletal2017b} estimate. Second, a correct modeling of the energy of the magnetar should include the explosion energy itself, $E_{\rm SN}$, as done by, e.g., \cite{KasenBildsten2010} and \cite{Kasenetal2016}. Nonetheless, the conclusion that many of the SLSNe cannot be exploded by neutrinos holds.

\cite{Thompsonetal2004} proposed that a rapidly rotating magnetized NS can blow a strong wind, termed a neutrino-magnetocentrifugally driven wind, and that this wind can account for hyperenergetic supernovae.  However, they require the magnetic field of the NS to be $\ga 10^{15} \G$, much larger than the values that \cite{Nicholletal2017b} deduce for the SLSNe studied here. Moreover, a wind will substantially spin-down the NS, such that the initial angular momentum is much larger than the  angular momentum of the magnetar. This requires very high specific angular momentum of the material that forms the NS, and a formation of an accretion disk is more likely even.

\begin{figure}
\begin{center}
\includegraphics[scale=0.62]{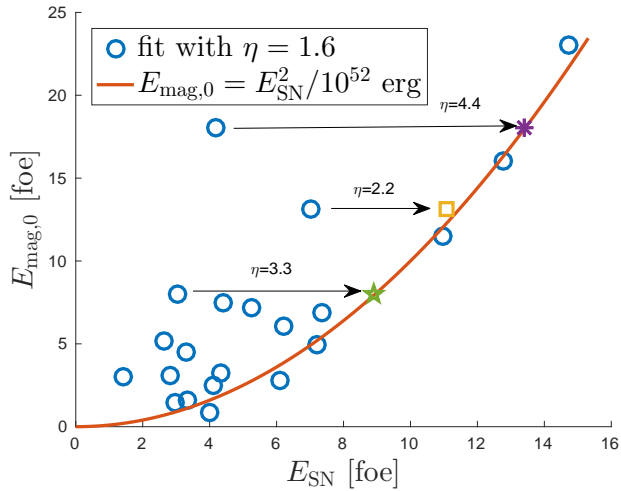}
\vskip +0.5 cm
\caption{A plane of the initial magnetar energy $E_{\rm mag,0}$ versus the explosion energy $E_{\rm SN}$. To calculate the explosion energy for the 21 systems that are placed on the graph we take here $\eta=1.6$ in equation (\ref{eq:Ek}). We also mark the consequence of taking higher values of $\eta>1.6$ for three systems. The line is a plot of the relation given in equation (\ref{eq:relation}) taken from \cite{Soker2017Mag2}.
}
\label{fig:relation}
\end{center}
\end{figure}

We are left then with jet-driven explosions. In an earlier paper on the relation between the JFM and magnetars \citep{Soker2017Mag2}, the following approximate relation between the initial energy of the magnetar and the explosion energy was derived
\begin{equation}
E_{\rm mag,0} \simeq \chi \frac{E^2_{\rm SN}}{10^{52} \erg},
\label{eq:relation}
\end{equation}
where $\chi \approx 1$ depends on the moment of inertia of the NS, the fraction of the gravitational energy of the accreted gas onto the NS that is carried by the jets, and the amount of angular momentum that is removed by the jets. The value of $\chi$ is expected to change somewhat from one system to another. Nonetheless, we plot this relation in Fig. \ref{fig:relation}, where we also place the 21 systems in a graph of the initial magnetar energy against the explosion energy. In calculating the explosion energy for this graph we take $\eta=1.6$ in equation (\ref{eq:Ek}). We also mark the consequence of taking higher values of $\eta$ for three systems. 

In discussing Fig. \ref{fig:relation} we should bear in mind the large uncertainties in the values of the kinetic energy of the SLSNe and the other magnetar parameters as derived by \cite{Nicholletal2017b}, and the somewhat different values that $\chi \approx 1$ is expected to take for different SLSNe in the JFM model \citep{Soker2017Mag2}. Despite these, it seems that the general behavior of the 21 SLSNe that are analyzed here is that the magnetar energy tends to increase with the explosion energy. This general behavior is the expectation of the JFM as depicted by the solid red line. 
For $\eta=1.6$ the predicted relation nicely bounds the SLSNe from the right and from below. It is possible that the kinetic energies of many of the SLSNe that are to the left of the line are more than a factor of 1.6 larger than the minimum kinetic energies that are derived by \cite{Nicholletal2017b}. 
The uncertainties in the values of $E_{\rm SN}$ resulting from the uncertainties in the magnetar parameters, that we estimate to have a typical value of about $\pm 50 \%$ before the larger uncertainties in the values of $\eta$ are considered, might somewhat displace individual objects on Fig. \ref{fig:relation}, but will not change the general trend.

\section{The r-process in the jet-driven explosion model}
\label{sec:rprocess}
 
The recently observed binary NS merger event GW170817 shows that r-process elements are formed in this process (e.g., \citealt{Metzger2017} for a summary of the event and references, and \citealt{Coteetal2018} for specific discussion of the r-process). However, it might be that another site is needed for the synthesis of r-process elements in low-metallicity stars made early in the evolution of the Galaxy (e.g., \citealt{Thielemannetal2017}). 

In principle, neutron-rich jets that are launched by the newly born NS in CCSNe might form r-process elements (e.g., \citealt{Winteleretal2012}).  
The problem with jets that are launched at several seconds from the formation of the NS is that because of the high flux of neutrinos that are emitted by the cooling NS, neutrons absorb electron-neutrinos and turn into protons. \cite{Fischeretal2010} and \cite{Hudepohletal2010} found that a neutrino driven wind becomes proton-rich by this process. The mass-loss rate in the neutrino-driven wind as presented by \cite{Hudepohletal2010} declines from about $0.03 M_\odot \s^{-1}$ to about $10^{-4} M_\odot \s^{-1}$ at $t=2\s$. 
The mass-loss rate in the jets of the jittering jets model is $\approx 10^{-2} M_\odot \s^{-1}$ \citep{PapishSoker2012, PapishSoker2014}, and it takes place in the first 2 seconds when the neutrino luminosity is very high. 

In any case, it is clear that if a large fraction of CCSNe are powered by jets, to be compatible with the low r-process abundance these jets cannot produce r-process elements. On average, there is an r-process mass of about $10^{-4} M_\odot$ per CCSN (e.g., \citealt{MathewsCowan1990, Thielemannetal2017}). Since a substantial amount is formed in a binary NS merger, the average mass of r-process elements in each CCSN should be less than about a few times $10^{-5} M_\odot$. 

In their simple spherically symmetric calculations \cite{PapishSoker2012} have found that the mass of the r-process elements that is formed in the jet-inflated bubbles of the jittering jets model is several times $10^{-4} M_\odot$. Namely, 10 times more than what is allowed by observations. 

As was already pointed out by \cite{Papishetal2015b}, there are several effects that are expected to substantially reduce the r-process elements mass as estimated by \cite{PapishSoker2012}. (1) At early times the jets are launched from a radius larger than the final radius of the NS. 
\cite{PapishSoker2012} and \cite{Papishetal2015b} pointed out that this leads to less neutron-rich matter in the jet.  
(2) The conversion of neutrons to protons by electron-neutrinos: The mass-loss rate of the jets is similar to that in the wind calculated by \cite{Hudepohletal2010} who found that the wind becomes proton-rich. \cite{Winteleretal2012} found a modest change in the neutron enrichment as a result of this process, but still one that can reduce the final production of r-process elements. 
(3) The third effect might turn out to be the most important one. In the jittering jets model the jets explode the star by interacting with the core material. The jets are shocked and form hot bubbles (so even if strong-r process elements have been synthesized in the jets, they will be disintegrated in the shock; \citealt{Papishetal2015b}). The final r-process elements are produced inside the hot bubbles. The simulations of such jets show that core material is mixed into the bubble \citep{PapishSoker2014b}. The mixing can take place as the jets drag gas from their surroundings and by instabilities that develop when the jets are shocked. This mixing is expected to further lower the mass of the r-process elements that are synthesized inside the hot bubble. 

Clearly an accurate calculation of the r-process in the jittering jets model is needed. 
At this stage we accept the conclusion of \cite{Papishetal2015b} that the average mass per CCSN event of r-process elements in the jittering jets model is very low, $\ll 10^{-4} M_\odot$. Namely, CCSNe cannot be even the rare site for synthesis of r-process elements in old low-metallicity stars. In rare cases jets that are formed at late times from fall-back material might form r-process elements as the NS is already cool and there is no core material anymore that the jets collide with. 

\cite{Papishetal2015b} suggested that the third possible site for r-process elements is a common envelope of a NS spiraling inside the envelope and core of a red supergiant.
The neutron star accretes mass and launches jets. This setting is different from jets in CCSNe in key ingredients \citep{Papishetal2015b}. (1) The old NS is cold and the neutrino flux is very low. This ensures that the neutron-rich gas that is launched from very close to the NS will stay so. (2) There is no dense core into which the jets are shocked. Hence, the r-process elements that are formed inside the jet do not disintegrate in a strong shock. 
(3) The NS ejects the massive envelope of the red supergiant, $\approx 10 -30 M_\odot$, but there is no iron production (unlike in a CCSN). This implies that if the giant is a very metal-poor star, i.e., this process takes place in the very young Galaxy, the abundance of r-process elements relative to iron can be large. Stars that are later formed from the ejected envelope will have very low iron abundances but will still have r-process elements.  
The binary NS merger site has a hard time to account for stars with low iron abundance but typical r-process abundance relative to iron (e.g., \citealt{Thielemannetal2017}). The NS common envelope r-process site might account for such stars. We reiterate the suggestion of \cite{Papishetal2015b} that a NS in a common envelope can form r-process elements, in particular in the metal-poor early Universe. 

\section{Summary}
\label{sec:summary}

In two previous papers one of us already argued that any CCSN that at late times is powered by a magnetar is expected to be exploded by jets \citep{Soker2016Mag, Soker2017Mag2}.
In the present paper we approached the question of the explosion mechanism from a different direction. We analyzed 21 out of the 38 SLSNe that their lightcurves were fitted with the magnetar model by \cite{Nicholletal2017b}, and calculated the explosion energy of these SLSNe. The rest of the SLSNe in the sample of \cite{Nicholletal2017b} have a magnetar spin-down time much shorter than the rise time to maximum light, and our analysis becomes less accurate; these require a self-consistent treatment of the explosion energy with the magnetar energy. 
We list our calculated explosion energies for three values of $\eta$, where $\eta$ is defined in equation (\ref{eq:Ek}), in the last three columns of Table \ref{Tab:Table1}.

In Fig. \ref{fig:relation} we place the 21 SLSNe on the plane of the initial magnetar energy $E_{\rm mag,0}$ versus the explosion energy $E_{\rm SN}$, where the explosion energies are calculated this time with $\eta=1.6$ for all SLSNe. We also mark the consequence of taking higher values of $\eta$ for three systems. We take the value of $\eta=1.6$ to show that the relation that was derived by \cite{Soker2017Mag2} that we plot by the solid red line (given here in equation \ref{eq:relation}), bounds the SLSNe from below and from the right. We speculate that for most of the SLSNe to the left of the line the kinetic energies were underestimated by a factor larger than $\eta=1.6$. Using larger values of $\eta$ would bring them toward the red line.  

Our main finding is that the explosion energies themselves of about half of the 38 SLSNe from the sample of \cite{Nicholletal2017b} are more than what the delayed neutrino mechanism can supply, $E_{\rm SN} > 2 \times 10^{51} \erg$. This has three implications. (1) The explosion cannot be driven by neutrinos. (2) The modeling of SLSNe with magnetars must include the explosion energy as a parameter, and cannot assume an explosion energy of $10^{51} \erg$. The explosion energy can be either a free parameter, or the relation  (\ref{eq:relation}) between the initial magnetar energy and the explosion energy can be used. (3) Jets can do more than drive the explosion on a time-scale of seconds. Jets might also power the ejecta at much later times, when some gas from the equatorial plane vicinity falls back and forms an accretion disk that launches late jets (e.g., \citealt{Gilkisetal2016}). Late powering by jets alongside the magnetar must be considered as well in fitting the light-curve.

Our group (e.g., \citealt{Papishetal2015a, Soker2017Mag2, GrichenerSoker2017}) has called for a paradigm shift from neutrino-driven explosions to jet-driven explosions of all CCSNe. Several recent studies that find links between the roles of jets in energetic explosions and in weaker explosions (e.g., \citealt{Marguttietal2014, Sobacchietal2017, Bearetal2017}) support our call (that was also echoed on weaker terms by \citealt{Piranetal2017}). The present study further strengthens our call for a paradigm shift toward jet-driven explosions of all CCSNe, under the condition that the mass of the r-process elements that are synthesized by the jets is very low. As we discuss in section \ref{sec:rprocess} this is likely to be the case. The jets most likely operate via a negative jet feedback explosion mechanism.

\section*{Acknowledgments}

We thank Brian Metzger for helpful comments. We thank an anonymous referee for promoting the discussion on the r-process (section 4). This research was supported by the E. and J. Bishop Research Fund at the Technion and by a grant from the Israel Science Foundation. A.G. is supported by the Blavatnik Family Foundation.

\end{document}